\documentclass[12pt]{article}
\usepackage{tocloft}
\usepackage{amsfonts}
\usepackage{amsmath}
\usepackage{amssymb}
\usepackage{array,slashed}
\usepackage{bigints}
\usepackage{bm}
\usepackage{booktabs}
\usepackage[nosort]{cite}
\usepackage{color}
\usepackage{dsfont}
\usepackage{float}
\usepackage{framed}
\usepackage{graphicx}
\usepackage{indentfirst}
\usepackage{mathrsfs}
\usepackage{multirow}
\usepackage{pdflscape}
\usepackage{tikz-cd}
\usepackage{setspace}
\usepackage{subdepth}
\usepackage{subfig}
\usepackage{titlesec}
\usepackage{wrapfig}
\usepackage[all]{xy}
\usepackage{young}
\usepackage[vcentermath]{youngtab}
\usepackage{relsize}
\usepackage{stackengine}
\usepackage{verbatim}
\usepackage{slashed}
\usepackage{color}

\usepackage{hyperref}
\hypersetup{colorlinks=true}
\hypersetup{linkcolor=black}
\hypersetup{citecolor=black}
\hypersetup{urlcolor=black}

\numberwithin{equation}{section}

\usepackage[left=2.5cm,right=2.5cm,top=2.5cm,bottom=3cm]{geometry}
\fontdimen2\font=1.2\fontdimen2{\jot}{5pt}

\titleformat{\section}{\large\bfseries}{\thesection.}{4pt}{}
\titlespacing{\section}{0pt}{20pt}{6pt}

\titleformat{\subsection}{\normalfont\bfseries}{\thesubsection.}{4pt}{}
\titlespacing{\subsection}{0pt}{15pt}{6pt}

\titleformat{\subsubsection}{\normalfont\itshape}{\thesubsubsection.}{4pt}{}
\titlespacing{\subsubsection}{0pt}{15pt}{6pt}

\titleformat{\paragraph}{\normalfont\itshape}{\theparagraph.}{4pt}{}
\titlespacing{\paragraph}{0pt}{15pt}{6pt}

\def\tilde{\widetilde}

\def\hat{\widehat}

\def\bar{\overline}

\DeclareMathAlphabet{\mathbfsf}{OT1}{cmss}{bx}{n}

\def\CZ{{\mathcal Z}}

\def\bZ{\mathbb{Z}}
\def\bR{\mathbb{R}}

\def\bQ{\mathbb{Q}}
\newcommand{\cA}{\mathcal{A}}

\newcommand{\cC}{\mathcal{C}}
\newcommand{\cD}{\mathcal{D}}

\newcommand{\cM}{\mathcal M}

\newcommand{\cO}{\mathcal{O}}

\newcommand{\cT}{\mathcal{T}}

\newcommand{\ed}{\,.}
\newcommand{\ec}{\,,}

\newcommand{\be}{\begin{equation}}
\newcommand{\ee}{\end{equation}}

\newcommand{\zero}{^{(0)}}

\newcommand{\one}{^{(1)}}
\newcommand{\two}{^{(2)}}
\newcommand{\three}{^{(3)}}
\newcommand{\four}{^{(4)}}
\newcommand{\five}{^{(5)}}

\DeclareFontShape{OT1}{cmr}{mx}{n}%
{<->cmr10}{}
\newcommand{\mytitlefont}{\fontseries{mx}\selectfont}
\DeclareMathAlphabet{\titlemath}{OT1}{cmr}{mx}{n}

\begin{document}
	
	

%
\begin{titlepage}
\begin{center}
~\\[1.5cm]
{\fontsize{27pt}{0pt} \mytitlefont 
The Non-Invertible Axial Symmetry\\ 
\vskip 10pt
 in QED Comes Full Circle}
~\\[1.25cm]
Adrien Arbalestrier, Riccardo Argurio, and Luigi Tizzano
~\\[0.5cm]
{~{\it Physique Th\'eorique et Math\'ematique and International Solvay Institutes\\
Universit\'e Libre de Bruxelles; C.P. 231, 1050 Brussels, Belgium}}

~\\[1.25cm]
			
\end{center}
\noindent
We revisit the possibility of constructing non-invertible topological defects for the axial symmetry of massless QED, despite its ABJ anomaly. Dressing the defects with a topological quantum field theory with mixed $U(1)$ and $\bR$-valued gauge fields, we are able to describe axial rotations of any rational or irrational angle. We confront our results with the existing proposals, in particular those that concern rational angles. We also provide the Symmetry TFT that reproduces the action of all such symmetry defects of QED. Finally, we discuss how similar techniques allow the study of condensation defects for a $\mathbb{R}$ global symmetry.

\vfill 
\begin{flushleft}
May 2024
\end{flushleft}
\end{titlepage}
%
		
	
\setcounter{tocdepth}{3}
\renewcommand{\cfttoctitlefont}{\large\bfseries}
\renewcommand{\cftsecaftersnum}{.}
\renewcommand{\cftsubsecaftersnum}{.}
\renewcommand{\cftsubsubsecaftersnum}{.}
\renewcommand{\cftdotsep}{6}
\renewcommand\contentsname{\centerline{Contents}}
	
\tableofcontents

\section{Introduction}
The study of symmetry constraints on the low-energy behavior of quantum field theories has a long and venerable history.
Thanks to the works of Adler \cite{Adler:1969gk} and Bell-Jackiw \cite{Bell:1969ts}, it is known that in theories with fermions
a continuous global symmetry can fail to persist at the quantum level. This phenomenon is commonly referred to as
an ABJ anomaly and has many interesting phenomenological consequences. 

At the same time, symmetries suffering from an ABJ anomaly should not be regarded as true global symmetries of a given quantum system; therefore, they seem completely useless for characterizing its RG flow.

Recently, this traditional viewpoint on the subject has been thoroughly reconsidered by the authors of \cite{Choi:2022jqy,Cordova:2022ieu}. They proposed that in $3+1d$ massless quantum electrodynamics (QED) a dense subset of the $U(1)_A$ axial symmetry group elements can be reinterpreted as non-invertible global symmetries.

The non-invertible nature of the axial symmetry is based on the modern understanding of global symmetry in quantum field theory which is itself rooted in the notion of topological symmetry defects \cite{Gaiotto:2014kfa}. Due to the ABJ anomaly, QED does not admit a gauge-invariant Noether current for the $U(1)_A$ symmetry. However, for rational values of the $U(1)_A$ rotation angle $\alpha = 2\pi p/N$ one can obtain a gauge-invariant topological symmetry defect $\cD^\bQ_{2\pi p/N}(\Pi\three)$ by dressing it with a $3d$ topological quantum field theory (TQFT). This TQFT is coupled to the $3+1d$ electromagnetic gauge field and only exists on the closed oriented 3-manifold $\Pi\three$. The topological defect $\cD^\bQ_{2\pi p/N}(\Pi\three)$ does not obey group multiplication law, in particular $\cD^\bQ_{2\pi p/N} \times \bar{\cD}^\bQ_{2\pi p/N} = \cC_{\bZ_N} \neq 1$ where $\cC_{\bZ_N}$ is known as the condensation defect \cite{Gaiotto:2019xmp} for the higher gauging \cite{Roumpedakis:2022aik} on $\Pi\three$ of a $\bZ\one_N$ subgroup of the $U(1)\one$ magnetic 1-form symmetry of QED. With the operators $\cD^\bQ_{2\pi p/N}$ we can therefore interpret the axial symmetry of QED as a novel kind of global symmetry which is non-invertible.

Although non-invertible axial rotations at rational angles already yield significant physical consequences  (see e.g.~\cite{Choi:2022jqy,Cordova:2022ieu}), this raises the question of why such a restriction is necessary. In this work, we take a further step in the study of the $U(1)_A$ axial symmetry in order to remove this restriction. 

We propose a novel topological symmetry defect, denoted by $\cD_\alpha(\Pi\three)$, which is gauge invariant for \emph{any} value of the rotation angle $\alpha \in [0,2\pi)$.\footnote{Alternative ideas have already appeared in the literature \cite{Karasik:2022kkq,GarciaEtxebarria:2022jky}. We will discuss the differences between these works and our approach at the end of Section \ref{NoninvU1}.} Even though $\alpha$ is now a continuous parameter, the symmetry defect $\cD_\alpha(\Pi\three)$ does not obey the group multiplication law since $\cD_\alpha\times\cD_{-\alpha} = \cC_{U(1)}\neq 1$. In this context $\cC_{U(1)}$ is interpreted as a condensation defect for the higher gauging on $\Pi\three$ (with discrete topology) of the \emph{full} $U(1)\one$  magnetic 1-form symmetry of QED.

The main technical tool that allows us to describe the $U(1)_A$ axial symmetry as a continuous non-invertible symmetry is a $3d$ TQFT of BF type based on two gauge fields taking values respectively in $U(1)$ and $\bR$ which we refer to as a mixed $U(1)/\bR$ BF theory. These kind of mixed BF theories have been recently studied in \cite{Antinucci:2024zjp} and \cite{Brennan:2024fgj}, in the context of formulating a symmetry TFT \cite{Gaiotto:2014kfa, Gaiotto:2020iye, Apruzzi:2021nmk, Freed:2022qnc, Kaidi:2022cpf, Antinucci:2022vyk, Bhardwaj:2023ayw, Bhardwaj:2023fca, Bhardwaj:2023bbf} for theories with $U(1)$ global symmetries.\footnote{Furthermore, see \cite{Apruzzi:2024htg} for an alternative approach aimed at replacing the Symmetry TFT with a non-topological bulk theory, dubbed Symmetry Theory. A Symmetry TFT for continuous non-Abelian global symmetries has been formulated in \cite{Antinucci:2024zjp,Brennan:2024fgj,Bonetti:2024cjk}. Finally, see \cite{Cvetic:2023plv} for a string theory analysis of $U(1)$ and $\cD^\bQ_{2\pi p/N}$ symmetry defects.} We explore the implications of our work for the symmetry TFT of QED, finding an interesting extension of the proposal of \cite{Antinucci:2024zjp}.

In addition, in Appendix \ref{minimalRsection} we present a different set of non-invertible topological defects dressed by a $3d$ Chern-Simons theory with gauge group $\bR$. These can be used to define condensation defects for an $\mathbb{R}$ global symmetry, which may be of independent interest for future studies.

Our approach can be further applied to a large class of $4d$ quantum field theories with a $U(1)$ global symmetry that also admit a more general anomaly structure that is typically associated with a $2$-group global symmetry \cite{Kapustin:2013uxa, Tachikawa:2017gyf, Cordova:2018cvg, Benini:2018reh}. In a separate publication \cite{Usinthefuture}, we analyze these examples and explain the main differences with QED, where this generalized structure is absent. This is not limited to four dimensions; in fact, a similar symmetry structure also appears in certain examples of $3d$ and $5d$ quantum field theories, as shown in \cite{Damia:2022rxw, Damia:2022bcd}, which we will also explore.

\section{Non-Invertible $U(1)_A$ Axial Symmetry}\label{NoninvU1}
In QED with massless fermions, the classical $U(1)_A$ axial symmetry
is broken by an ABJ anomaly. The non-conservation law of the axial current is given by
\be\label{ABJ}
d* j\one_A=\frac{1}{4\pi^2}dA\one\wedge dA\one\ec
\ee
where $A\one$ denotes the $U(1)$ dynamical electromagnetic gauge field of QED.\footnote{Throughout this work, depending on the context, for a given object $\cO$ the notation $\cO^{(p)}$ will be used to denote either a $p$-form, a $p$-dimensional manifold or a $p$-dimensional homology cycle.} From this relation, we can define topological operators implementing transformations of the $U(1)_A$ axial symmetry:
\be \label{U(1)A operator}
U_{\alpha}(\Pi\three)=\exp\left(i\frac{\alpha}{2}\int_{\Pi\three}* j\one_A-\frac{1}{4\pi^2}A\one\wedge dA\one\right)\ec
\ee
where $\Pi\three$ is a closed 3-manifold and $\alpha \in [0,2\pi)$. These operators are however not invariant under large gauge transformations of $A\one$ for general values of $\alpha$.\footnote{We will be adopt the convention of \cite{Choi:2022jqy} in which the axial symmetry $U(1)_A$ acts as 
$\Psi \to e^{i\gamma_5\alpha/2 }\Psi$.
Note that $\alpha=2\pi$ implements a transformation that is part of the $U(1)$ gauge group of QED.} This issue can be remedied by dressing the naive axial symmetry operator with a 3d TQFT:
\be\label{newD}
\cD_{\alpha}(\Pi\three) = \exp\left(i\int_{\Pi\three}\frac{\alpha}{2} * j\one_A + \frac{i}{2\pi}\int_{\Pi\three}-\frac{\alpha}{4\pi}c\one\wedge dc\one+\Phi\one\wedge(dc\one+dA\one)\right)\ec
\ee
where $c\one$ and $\Phi\one$ are respectively dynamical $\bR$-valued and $U(1)$-valued 1-form gauge fields that only live on $\Pi\three$ (here and below, we sometimes omit to write explicitly the path integral over the fields localized on the defect). The above TQFT is a three-dimensional example of the mixed $U(1)/\mathbb{R}$ BF theory recently studied in \cite{Antinucci:2024zjp}, coupled to the 4d bulk QED gauge field $A\one$.\footnote{The standard analysis of BF theory with $U(1)$ gauge fields can be found in \cite{Maldacena:2001ss,Banks:2010zn,Kapustin:2014gua}.} In what follows, we will denote this theory by:
\be\label{Talpha}
\cT^\alpha[dA\one] = \frac{i}{2\pi}\int_{\Pi\three}-\frac{\alpha}{4\pi}c\one\wedge dc\one+\Phi\one\wedge(dc\one+dA\one)\ed
\ee
Note that, for $\alpha \in 2\pi\bQ$, the above proposal is very close in spirit to the discussion of \cite{Choi:2022jqy, Cordova:2022ieu}. We will highlight all the technical differences with the constructions of those papers in section \ref{condensection}. More details on the $\cT^\alpha$ theory, including alternative formulations, are discussed in the Appendices.

The operator \eqref{newD} can thus be written more compactly as
\be
\cD_{\alpha}(\Pi\three)= \exp\left(i\int_{\Pi\three}\frac{\alpha}{2}*j\one_A + \cT^\alpha[dA\one]\right)\ed
\ee
A different perspective on the theory \eqref{Talpha} and the associated topological defect \eqref{newD} can be developed by employing the approach of gauging a global symmetry in half-spacetime. For finite global symmetry groups this procedure was first introduced in \cite{Roumpedakis:2022aik} and advocated in this context in \cite{Choi:2022jqy}. Here we present a generalization of this idea to accommodate a $U(1)$ symmetry group.

Let us consider the following action defined on a 4-dimensional manifold $\Omega\four$:
\be\label{bulk Talpha}
S_4[dA\one]=-\frac{i}{2\pi}\int_{\Omega\four} B\two\wedge dc\one+ B\two\wedge dA\one +\phi\two\wedge d\Phi\one -\frac{\alpha}{4\pi}\phi\two\wedge \phi\two + \phi\two\wedge B\two\ec 
\ee
where $B\two$ and $\Phi\one$ are $U(1)$-valued fields while $c\one$ and $\phi\two$ are $\bR$-valued. 
The action \eqref{bulk Talpha} can be interpreted as the result of performing two successive topological gaugings. The first one is the gauging (with discrete topology) of the magnetic $U(1)\one$ symmetry with current $*j\two=\frac{1}{2\pi}dA\one$, where the flatness of $B\two$ is enforced by the $\mathbb{R}$-valued field $c\one$ appearing in the first term. The second gauging concerns the Pontryagin dual symmetry $\bZ\one=\hat{U(1)}\one$ and involves the other $\mathbb{R}$-valued gauge field $\phi\two$, whose flatness is enforced by the BF coupling to $\Phi\one$. Moreover, one crucially adds a counterterm proportional to $\alpha$.\footnote{Indeed, this procedure is similar to the $CST^{-p}S$ gauging in \cite{Choi:2022jqy}, in that context the two gauged symmetries are both $\bZ_N\one$ symmetries.}

When $\Omega\four$ is closed, the action \eqref{bulk Talpha} is invariant under the following gauge transformations:
\begin{equation}
    \begin{split}
        &B\two\rightarrow B\two+d\Lambda\one_B\ec \quad \phi\two\rightarrow \phi\two +d\lambda\one_{\phi}\ec\\
        &c\one\rightarrow c\one+d\lambda\zero_c-\lambda\one_{\phi}\ec \quad \Phi\one\rightarrow \Phi\one +d\Lambda\zero_{\Phi}-\Lambda\one_B+\frac{\alpha}{2\pi}\lambda\one_{\phi}\ed
    \end{split}
\end{equation}
To make contact with the previous analysis we now assume that $\Omega\four$ has a boundary such that $\partial \Omega\four=\Pi\three$. Furthermore, we consider Dirichlet boundary conditions on $B\two$ and $\phi\two$ which restrict the gauge transformations $\Lambda_B\one$ and $\lambda_{\phi}\one$ to vanish at the boundary of $\Omega\four$.

We can straightforwardly integrate out $\phi\two$ and $B\two$ from \eqref{bulk Talpha}. After redefining $\Phi\one\rightarrow \Phi\one-\frac{\alpha}{2\pi} c\one$ and reducing to the boundary, we obtain
\begin{equation}\label{4d Talpha anomaly}
    \begin{split}
        S_4[dA\one]&=\frac{i}{2\pi}\int_{\partial \Omega\four} \left(-\frac{\alpha}{4\pi}c\one\wedge dc\one+\Phi\one\wedge(dc\one+dA\one)\right)+\frac{i\alpha}{
        8\pi^2}\int_{\partial \Omega\four} A\one\wedge dA\one\ec
    \end{split}
\end{equation}
which is what we expect in order to reproduce the bulk ABJ anomaly \eqref{ABJ}. 

The virtue of the above approach is that it provides a detailed verification of the conservation, or more broadly, the topological property of $\cD_{\alpha}$, defined in \eqref{newD}.

\subsection{Action on Operators}\label{actionop}
As mentioned in the previous section, QED has a magnetic $1$-form symmetry $U(1)\one$ 
whose codimension $2$ topological symmetry defect is $U_{\theta}(\Sigma\two) = \exp(\frac{i\theta}{2\pi}\int_{\Sigma\two}dA\one)$. The associated
charged objects are 't Hooft lines of charge $q\in \bZ$ which we denote by $H_q(\gamma\one)$. Since the non-invertible topological
defect $\cD_{\alpha}$ is constructed by gauging the $U(1)\one$ symmetry in half of the spacetime, it is important to determine 
how it acts on 't Hooft lines.

The equations of motion of the $\cT^{\alpha}$ theory can be written as
\begin{equation}
    dc\one+dA\one=0\ec\quad d\Phi\one+\frac{\alpha}{2\pi}dA\one=0\ed
\end{equation}
The first equation above implies that any $U(1)\one$ magnetic symmetry operator can end topologically on $\cT^{\alpha}$ by opening on a $c\one$ line. It also implies that $dA\one$ must have trivial fluxes on $\Pi\three$: 
\be\label{trivial}
\int_{\Sigma\two}dA\one=0\ec
\ee 
where $\Sigma\two \subset \Pi\three$ is a closed surface. This annihilates magnetic surface operators. 

However, it is possible to obtain something more interesting by inserting an operator 
\be\label{insertion}
\exp\left(iq\int_{\gamma\one}\Phi\one+i\frac{q\alpha}{2\pi}\int_{\Sigma\two_{\gamma}}dA\one\right)\ec
\ee 
where $\Sigma\two_{\gamma}$ is an open surface with 1d boundary $\gamma\one\subset \Pi\three$.
This modifies the equation of motion to: 
\be\label{modEOM}
dc\one+dA\one=-2\pi q \delta\two(\gamma)\ec
\ee
where $\delta\two(\gamma)$ is the 2-form Poincar\'e dual to $\gamma\one$ within $\Pi\three$. The equation \eqref{trivial} is thus modified to 
\be\label{nontrivial flux}
\int_{\Sigma\two}dA\one=-2\pi q I(\Sigma\two,\gamma\one)\ec
\ee 
where $I(\Sigma,\gamma)$ is the intersection number of $\Sigma\two$ and $\gamma\one$. 

To summarize, when a non-invertible symmetry operator $\cD_{\alpha}(\Pi\three)$ wraps a magnetic 't Hooft line of charge $q$, the resulting correlation function is trivial unless we insert an operator \eqref{insertion}. In turn, the operator \eqref{insertion} is gauge invariant if and only if $q\in \bZ$ implying that the flux of $dA\one$ is quantized on  surfaces of $\Pi\three$. Indeed, this is perfectly consistent with the fact that $A\one$ is a $U(1)$-valued 1-form gauge field.

\begin{figure}
    \centering
\includegraphics[width=0.45\linewidth]{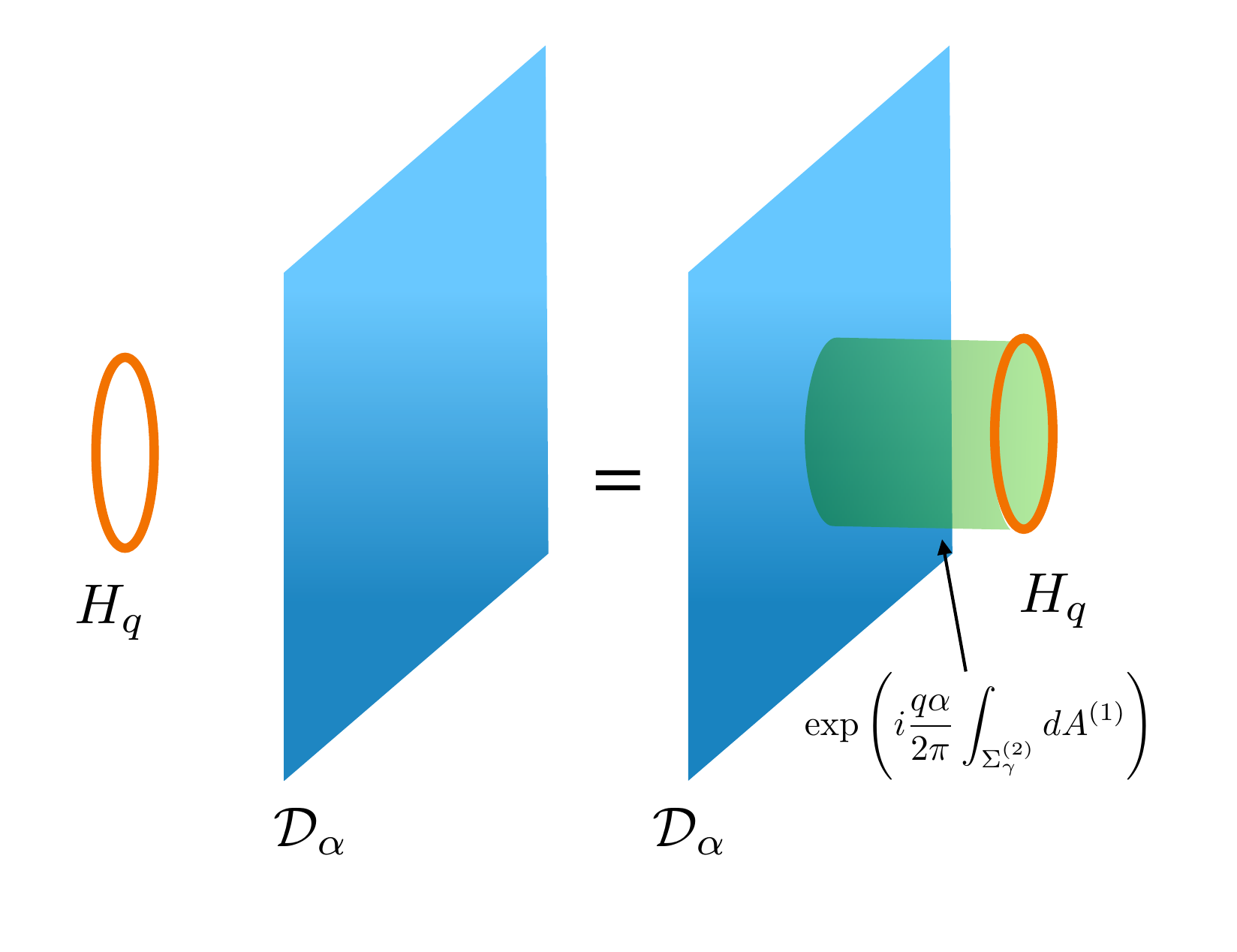}
   \caption{Action of the non-invertible symmetry defect $\cD_{\alpha}$ when it is swept across a magnetic 't Hooft operator $H_q$.}
   \label{fig1}
\end{figure}

Finally, as we sweep a $\cD_{\alpha}(\Pi\three)$ operator across a magnetic 't Hooft line $H_q$ (see figure \ref{fig1}) it is easy to deduce from \eqref{modEOM} that $H_q$ becomes a non-genuine line operator attached to a magnetic symmetry surface:
\be\label{nongenHq}
H_q\exp\left(i\frac{q\alpha}{2\pi}\int_{\Sigma_{\gamma}\two}dA\one\right)\ed
\ee
This effect is completely analogous to what has been observed in \cite{Choi:2022jqy, Cordova:2022ieu} for rotations $\alpha \in 2\pi \mathbb{Q}$. In our setup though, we can consistently define non-invertible symmetry defects for generic irrational angles $\alpha\in 2\pi (\bR\!\setminus\!\mathbb{Q})$ and therefore {\em all} 't Hooft lines become non-genuine when acted upon by such symmetry defects. 

\subsection{Parallel Fusion and Condensation Defects}\label{condensection}
We would now like to study the properties of the topological operator \eqref{newD} under parallel fusion. Let us first consider the parallel fusion of $\cD_{\alpha}(\Pi\three)$ with $\cD_{-\alpha}(\Pi\three)$, namely the defect implementing the opposite axial rotation, which is given by:\footnote{We can alternatively see $\cD_{-\alpha}(\Pi\three)$ as $\cD_\alpha(\bar\Pi\three)$, where $\bar\Pi\three$ is the orientation reversal of the three-manifold $\Pi\three$. Their respective expressions are strictly equivalent as functionals of $A\one$.}
\be
\begin{split}
    \cD_{\alpha}(\Pi\three)\times& \cD_{-\alpha}(\Pi\three)=\int D[c\one,\Phi\one,\overline{c}\one,\overline{\Phi}\one]\exp\left(\frac{i}{2\pi}\int_{\Pi\three}-\frac{\alpha}{4\pi}c\one\wedge dc\one\right.\\
    &\left. +\Phi\one\wedge(dc\one+dA\one)+\frac{\alpha}{4\pi}\overline{c}\one\wedge d\overline{c}\one+\overline{\Phi}\one\wedge( d\overline{c}\one+dA\one)\right)\ed
\end{split}
\ee
After performing the following field redefinitions
\be
\Phi\one\rightarrow\Phi\one -\overline{\Phi}\one\ec \quad \overline{c}\one\rightarrow \overline{c}\one+c\one \ec\quad \overline{\Phi}\one\rightarrow \overline{\Phi}\one-\frac{\alpha}{2\pi} c\one-\frac{\alpha}{4\pi}\overline{c}\one\ec
\ee
we are left with:
\be\label{cond}
\cD_{\alpha}(\Pi\three)\times \cD_{-\alpha}(\Pi\three)= {\cal Z}_{U(1)/\mathbb{R}} \times \int D[c\one,\Phi\one]\exp\left(\frac{i}{2\pi}\int_{\Pi\three}\Phi\one\wedge(dc\one+dA\one)\right)\ed
\ee
The normalization factor ${\cal Z}_{U(1)/\mathbb{R}}$ is expressed in terms of the following path integral:
\be\label{normalization}
\mathcal{Z}_{U(1)/\bR} =\int D[\overline{c}\one,\overline{\Phi}\one]\exp\left(\frac{i}{2\pi}\int_{\Pi\three}\overline{\Phi}\one \wedge d\overline{c}\one\right) \ed
\ee
This is the partition function of the mixed $U(1)/\bR$ BF theory on a closed 3-manifold $\Pi\three$. Since in our case this theory features an $\bR$-valued gauge field the final expression will be formally infinite. In this work, we will not need to evaluate the path integral \eqref{normalization}. The quantum mechanical treatment of mixed $U(1)/\bR$ BF theories is an interesting problem on its own which should be treated separately. For this reason, we will limit ourselves to keeping this normalization factor explicit in the relevant expressions.

Let us now examine in more detail the remaining term in \eqref{cond}, which defines the condensation defect. Due to the coupling to $dA\one$, we first integrate out $c\one$. This reduces the path integral of $\Phi\one$ to be over $H^1(\Pi\three, U(1))$. By the universal coefficient theorem (assuming that there is no torsion), we have that $H^1(\Pi\three, U(1)) \cong U(1)^{b_1(\Pi\three)}$ where $b_1(\Pi\three)$ is the first Betti number of $\Pi\three$. The final result can be expressed as a sum over insertions of magnetic 1-form symmetry defects on all non-trivial 2-cycles $\gamma\two_i$ of $\Pi\three$ and of all angles $\theta_i$, which parameterize the holonomies of $\Phi\one$:
\be\label{condU1}
\cC_{U(1)}(\Pi\three) = \prod_{\gamma_i\two \in X}\left[\int_{0}^{2\pi}d\theta_i \exp\left(\frac{i\theta_i}{2\pi}\int_{\gamma_i\two}dA\one\right)\right]\ec
\ee
where $X$ is a minimal generating set of $H_2(\Pi\three,\bZ)$, with $\vert X\vert=b_2(\Pi\three)=b_1(\Pi\three)$, and we have chosen a convenient normalization. For instance, when $\Pi\three$ is topologically $S^1\times S^2$, we would simply get:
\be\label{condU1simple}
\cC_{U(1)}(S^1\times S^2) = \int_0^{2\pi}d\theta \exp\left(\frac{i\theta}{2\pi}\int_{S^2}dA\one\right)\ed
\ee

In summary, the parallel fusion of two symmetry defects \eqref{cond} gives rise to a condensation defect for a $U(1)$ symmetry:
\be\label{condensation U(1)U(1)}
    {\cD}_{\alpha}(\Pi\three)\times {\cD}_{-\alpha}(\Pi\three)={(\mathcal{Z}_{U(1)/\bR})}\ \cC_{U(1)}(\Pi\three)\ed
\ee
This signifies that the full axial symmetry group $U(1)_A$ is a non-invertible global symmetry of massless QED. 

Comparing the defect $\cC_{U(1)}(\Pi\three)$ with $\cD_{\alpha}(\Pi\three)$, as defined in section \ref{NoninvU1}, it is evident that $\cC_{U(1)}(\Pi\three)=\cD_0(\Pi\three)$. The action of $\cC_{U(1)}(\Pi\three)$ on magnetic 't Hooft lines can thus be deduced from $\cD_{\alpha}(\Pi\three)$ by setting $\alpha=0$. From the discussion in section \ref{actionop}, when $\cD_{\alpha}(\Pi\three)$ is swept across a magnetic line $H_q$ an open magnetic surface operator $\exp\left(iq\int_{\gamma\one} \Phi\one+i\frac{q\alpha}{2\pi}\int_{\Sigma\two}dA\one\right)$ is generated. The first term is present also when $\alpha=0$, and it is required to have the non trivial flux quantization \eqref{nontrivial flux}. 
This means that when $\cC_{U(1)}$ is swept across $H_q$, it leaves it a genuine operator, but a line operator is generated on $\cC_{U(1)}$. In other words, in a configuration where $\cC_{U(1)}(\Pi\three)$ wraps $H_q$, the correlation function can only be non-trivial if an operator $\exp\left(iq\int_{\gamma\one} \Phi\one\right)$ is inserted on $\Pi\three$. In particular, if the $\cC_{U(1)}$ that wraps $H_q$ does {\em not} contain the suitable line, it annihilates it upon shrinking. This establishes the non-invertible action of the $U(1)$ condensation defect $\cC_{U(1)}(\Pi\three)$ on $H_q$.

We can similarly study the fusion of two defects with arbitrary angles, and we find
\be
\cD_{\alpha}(\Pi\three)\times \cD_{\beta}(\Pi\three)=(\mathcal{Z}_{U(1)/\bR})\ \cD_{\alpha+\beta}(\Pi\three)\ed
\ee
This expression is derived in Appendix \ref{fusionsection}. Note that the normalization factor appearing above is necessary to ensure that the right and left sides have the same action on local operators. Indeed, when $\Pi\three = S^3$, the partition function $\CZ_{U(1)/\bR}$ coincides with the quantum dimension of  $\cD_\alpha$.

We would now like to highlight the differences between this study and the works \cite{Choi:2022jqy, Cordova:2022ieu}, which first revealed the non-invertible nature of axial symmetry in QED. A key distinction of the approach outlined in this paper from those in previous studies is that the construction presented here, involving the mixed $U(1)/\bR$ BF theory $\cT^\alpha$, gives rise to a non-invertible symmetry defect for \emph{any} value of the $U(1)_A$ rotation angle $\alpha \in [0,2\pi)$. Instead, in the approach of \cite{Choi:2022jqy, Cordova:2022ieu} the angle $\alpha$ is restricted to \emph{rational} values $\alpha = 2\pi p/N \in 2\pi \bQ$, $\gcd(p,N)=1$:
\be\label{Drational}
\cD_{2\pi p/N}^{\bQ}(\Pi\three) =  \exp\left(i\int_{\Pi\three}\frac{2\pi p}{2N}*j\one_A + \cA^{N,p}\left[dA\one/N\right]\right)\ec
\ee
where $\cA^{N,p}$ is the minimal 3d TQFT with 1-form symmetry $\bZ\one_N$ and anomaly $p \simeq p + N$ \cite{Hsin:2018vcg}.\footnote{See Appendix \ref{fusionsection} for additional details on the minimal  $\cA^{N,p}$ 3d TQFT.} As can be seen from the above formula, the TQFT $\cA^{N,p}[dA\one/N]$ is coupled to a background for a $\bZ\one_N$ subgroup of the $U(1)\one$ of the magnetic one-form symmetry. The parallel fusion rule of two defects \eqref{Drational} can be expressed as:\footnote{We again adopt a convenient normalization and tacitly assume the presence of torsion factors when needed.}
\be
\begin{split}
\cD^\bQ_{2\pi p/N}(\Pi\three) \times \bar{\cD}^\bQ_{-2\pi p/N}(\Pi\three) = \cC_{\bZ_N}(\Pi\three) 
&=\sum_{\gamma\two \in H_2(\Pi\three, \bZ_N)} \exp\left(\frac{i}{N}\int_{\gamma\two}dA\one\right)\\
&= \prod_{\gamma_i\two \in X} \sum_{p_i=0}^{N-1} \exp\left(i\frac{p_i}{N}\int_{\gamma_i\two}dA\one\right)\ec
\end{split}
\ee
where in the second line we made it more similar to the expression in \eqref{condU1}.
This is interpreted as higher-gauging of a $\bZ\one_N$ subgroup of the $U(1)\one$ magnetic symmetry of QED. It differs from the approach of this note, where the symmetry defect \eqref{newD} enables us to higher-gauge (with discrete topology) the entire $U(1)\one$ magnetic symmetry group as in formula \eqref{condU1}.

Using the results from appendix \ref{fusionsection}, it is possible to show that:
\be
\cD_{\alpha}(\Pi\three) \times \cD^\bQ_{2\pi p/N}(\Pi\three) = (\CZ_{\cA^{N,p}})\ \cD_{\alpha+ 2\pi p/N}(\Pi\three)\ec
\ee
where $\CZ_{\cA^{N,p}}$ is the partition function of the minimal 3d TQFT on $\Pi\three$. Therefore, we can always absorb a $\cD^\bQ_{2\pi p/N}(\Pi\three)$ into the defect \eqref{newD}. From this relation we also deduce that, since $\cD_0(\Pi\three)=\cC_{U(1)}(\Pi\three)$, we have: 
\be
\cD_{2\pi p/N}(\Pi\three)=\frac{1}{\CZ_{\cA^{N,p}}}\,\cC_{U(1)}(\Pi\three)\times \cD^{\mathbb{Q}}_{2\pi p/N}(\Pi\three)\ed
\ee
It is then clear that the defects $\cD_{2\pi p/N}$ are less minimal than the $\cD^\bQ_{2\pi p/N}$ ones. However, they yield a more unified treatment for all angles. In particular,
recall that when $\cC_{U(1)}$ crosses a charge $q$ magnetic 't Hooft line $H_q$, $H_q$ remains a genuine operator. As a result, the non-invertible symmetry action implemented by
$\cD_{\alpha}$ on $H_q$ is the same as the one of $\cD^{\bQ}_{2\pi p/N}$. However, the junction operator living on $\cD^{\bQ}_{2\pi p/N}$ is not equivalent to the junction operator living on $\cD_{2\pi p/N}$. In particular, when $q=N$, the junction operator living on $\cD^{\bQ}_{2\pi p/N}$ is trivial  but the one living on $\cD_{2\pi p/N}$ is not. This is actually a welcome feature, since it removes the subtleties related to the behavior of junctions under fusion of defects, as studied in \cite{Copetti:2023mcq}. 

Finally, we comment on a completely different approach explored in \cite{Karasik:2022kkq,GarciaEtxebarria:2022jky} aimed at restoring the full $U(1)_A$ axial symmetry with non-invertible symmetry defects. In those proposals, the naive axial symmetry operator \eqref{U(1)A operator} is made gauge invariant by essentially path integrating over its gauge orbit. This enforces the constraint \eqref{trivial}, which, however, cannot be relaxed by introducing additional operators on $\Pi\three$ as we did in equation \eqref{insertion}. Furthermore, one can compute the condensation defect derived from the symmetry operators in \cite{Karasik:2022kkq,GarciaEtxebarria:2022jky} to realize that these are {\em not} condensations of the full $U(1)\one$ magnetic symmetry, as in our proposal.

\section{Symmetry TFT Analysis}\label{symTFTanalysis}
As we have seen in section \ref{NoninvU1}, using the 3d TQFT $\cT^\alpha$ we can dress the naive $U(1)_A$ axial symmetry defect of QED and restore a full-fledged $U(1)$ non-invertible global symmetry. Given this novel observation, it is interesting to revisit how this idea fits in the recent proposal \cite{Antinucci:2024zjp} describing the symmetry TFT of QED. 

Let us first recall that this symmetry TFT has a  five-dimensional action given by:
\be\label{symQED}
    S_5=\frac{i}{2\pi}\int_{X\five} -b\three \wedge dA\one-f\two \wedge dC\two +\frac{1}{2\pi} A\one\wedge f\two\wedge f\two + \frac{1}{6\pi} A\one\wedge dA\one\wedge dA\one\ec
\ee
where $b\three$, $f\two$ are $\bR$-valued gauge fields while $A\one$, $C\two$ are $U(1)$-valued. In QED both cubic anomaly coefficients are equal to 2. The action is invariant under the following gauge transformations:
\begin{equation}\label{gaugeTrans4dTFT}
\begin{split}
    &A\one\rightarrow A\one+d\lambda_A\zero\ec\qquad\qquad
    f\two\rightarrow f\two+d\lambda_f\one\ec\\
    &C\two\rightarrow C\two+d\lambda_C\one+\frac{1}{\pi}\lambda_A\zero (f\two+d\lambda_f\one)+\frac{1}{\pi}\lambda_f\one\wedge A\one\ec\\
    &b\three\rightarrow b\three+d\lambda_b\two+\frac{1}{\pi}f\two\wedge \lambda_f\one+\frac{1}{2\pi}d\lambda_f\one\wedge \lambda_f\one\ec
\end{split}
\end{equation}
and the associated equations of motion are:
\begin{equation}
dA\one=0\ec\quad df\two=0\ec\quad dC\two-\frac{1}{\pi}f\two\wedge A\one=0\ec \quad
db\three-\frac{1}{2\pi}f\two\wedge f\two=0\ed
\end{equation}
The genuine gauge invariant topological operators of the theory are given by:
\be\label{U,W}
W_n(\gamma\one)=\exp\left(in\int_{\gamma\one} A\one\right) \ec \qquad U_\beta(\Sigma\two)=\exp\left(\frac{i\beta}{2\pi}\int_{\Sigma\two} f\two\right)\ec
\ee
with $\beta\in [0,2\pi)$. At the same time, we can also construct non-genuine topological operators in the following way:
\begin{align}
T_m(\Sigma\two,\Omega\three)&=\exp\left(im\int_{\Sigma\two} C\two-\frac{im}{\pi}\int_{\Omega\three} A\one\wedge f\two\right)\ec\label{nongenT}\\
V_\alpha(\Pi\three,\Omega\four)&=\exp\left(\frac{i\alpha}{4\pi}\int_{\Pi\three} b\three-\frac{i\alpha }{8\pi^2}\int_{\Omega\four}  f\two\wedge f\two\right)\ed\label{nongenV}
\end{align}
where $\partial\Omega\three = \Sigma\two$ and $\partial\Omega\four = \Pi\three$, and $\alpha\in[0,4\pi)$.\footnote{For $\alpha=2\pi$, it can be shown that the operator \eqref{nongenV} does not depend on $\Omega\four$. In particular, when $\Omega\four$ is closed, the second term in \eqref{nongenV} is in $2\pi\bZ$  since $\int f\two\in 2\pi\bZ$ and we assume that $\Omega\four$ is spin. Thus $V_{2\pi}(\Pi\three)$ is a genuine operator.} The above operators follow the standard braiding (as discussed e.g.~in \cite{Antinucci:2024zjp}):
\begin{align}\label{5d WV}
&\left\langle W_n(\gamma\one) V_\alpha(\Pi\three)\right\rangle=\exp\left(\frac{in\alpha}{2} \,\text{Link}(\gamma\one,\Pi\three)\right)\ec\\
&\left\langle T_m(\Sigma\two_1) U_\beta(\Sigma\two_2)\right\rangle=\exp\left(im\beta\,\mathrm{Link}(\Sigma\two_1,\Sigma\two_2)\right)\ed
\end{align}
More interestingly, they also display additional less standard relations due to the cubic terms in the action \eqref{symQED}:
\begin{align} \label{5d VT}
    &\left\langle V_\alpha(\Pi_1\three,\Omega_1\four)T_m(\Sigma_2\two,\Omega_2\three)\right\rangle=\left\langle V_\alpha(\tilde{\Pi}_1\three,\tilde{\Omega}_1\four)T_m(\Sigma_2\two,\Omega_2\three)U_{m\alpha}(\tilde{\Omega}_1\four\cap \Omega_2\three)\right\rangle\ec\\
    \label{5d TT}
    &\left\langle T_n(\Sigma_1\two,\Omega\three_1)T_m(\Sigma_2\two,\Omega\three_2)\right\rangle=\left\langle T_n(\tilde{\Sigma}_1\two,\tilde{\Omega}_1\three)T_m(\Sigma_2\two,\Omega_2\three)W_{2nm}(\tilde{\Omega}_1\three\cap \Omega_2\three)\right\rangle\ed
\end{align}
where, on the left hand side, we assumed that $\Omega_1\cap \Omega_2=0$. On the right-hand side, we used a superscript $\sim$ to describe manifolds obtained from a continuous deformation of the corresponding manifolds on the left-hand side. The genuine operators $U_{ m\alpha}$ and $W_{2 nm}$ are typically defined on open manifolds terminating on $T_m$ and $V_\alpha$.

Note that this kind of relations are similar to those appearing among defects participating in a higher-group symmetry structure \cite{Hidaka:2020iaz,Hidaka:2020izy,Brennan:2020ehu,Choi:2022fgx}. However, we emphasize that in QED such a structure is not present.

Let us now see how this 5d theory can capture the existence of a continuous family of non-invertible axial symmetry defects in the 4d theory. The relations discussed above, including \eqref{5d VT}, will be essential. Nevertheless, it will be necessary to have a genuine version of all bulk operators.

\subsection{Boundary Conditions and Genuine Bulk Operators}
In the standard symmetry TFT paradigm, the list of topological operators \eqref{U,W}, \eqref{nongenT}, \eqref{nongenV}, supplemented by appropriate boundary conditions on the dynamical 5d bulk fields will give rise to both topological symmetry operators and non-topological charged operators of the boundary QED theory. Our goal in this section is precisely to describe such procedure. In what follows, we will always consider the 5d Symmetry TFT on a five-manifold $X\five$ with two distinguished boundaries: one is the boundary where the physical theory lives, denoted by $\cM_P$, while the second boundary, denoted by $\cM$, will be used to discuss how to impose boundary conditions for the bulk dynamical fields.

The physical theory has local operators charged under the $U(1)_A$ symmetry and line operators charged under the magnetic $U(1)\one$ symmetry. From the Symmetry TFT perspective, these operators are respectively located at the boundary of open lines $W_n$ and open surfaces $T_m$ ending on $\cM_P$. The operator $V_{2\pi}$ is the bulk operator implementing an axial symmetry transformation of parameter $2\pi$ in $\cM_P$. This corresponds to a gauge transformation and the boundary conditions on $\cM$ must trivialize $V_{2\pi}$. Since the charged operators of the physical theory are genuine, we must choose boundary conditions on $\cM$ allowing $W_n$ and $T_m$ to terminate on this boundary. Consider first $W_n$. Due to its non-trivial braiding with $V_{2\pi}$ in \eqref{5d WV}, $W_n$ can terminate on $\cM$ only if $n\in 2\bZ$. This means $W_1$ does not terminate on a gauge invariant operator of $\cM_P$. Indeed, local gauge invariant operators can only carry an even charge under axial symmetry, as they are powers of a fermion bilinear. For the boundary conditions of $T_m$, we encounter two potential problems: the operators $T_m$ are not genuine, as indicated in \eqref{nongenT}, and moreover they have non-trivial relations given in \eqref{5d TT}. The second issue is easily fixed if we also trivialize $W_{2n}$ on the topological boundary, which is what we set out to do.\footnote{As often occurs when anomaly terms are present in the bulk, the choice of boundary conditions is restricted \cite{Kaidi:2023maf,Zhang:2023wlu, Antinucci:2023ezl, Cordova:2023bja, Argurio:2024oym}.} We will then impose Dirichlet boundary conditions on $\cM$ for the $U(1)$ fields $A\one$ and $C\two$.\footnote{Strictly speaking, $A\one$ should have a boundary condition that still leaves free its $\bZ_2$ part, so that $W_1$ does not end on $\cM$. In a complementary fashion, $V_{2\pi}$ is trivialized on the boundary if we impose Dirichlet boundary conditions on the $\bZ_2$ part of $b\three$. In the following, we will neglect this subtlety to avoid clutter.}

The first issue mentioned above is more serious: only a genuine operator can terminate trivially on $\mathcal{M}$ and $\cM_P$. To recover magnetic 't Hooft lines in the physical theory we need to consider the topological operator $T_m(\Sigma\two,\Omega\three)$ and remove its dependence on the open surface $\Omega\three$. We thus define a new genuine topological operator obtained by stacking with a 2d TQFT:
\be\label{genT}
T_m(\Sigma\two)=\exp\left(im\int_{\Sigma\two} C\two +\cT_{2d}^{2m}[A\one, f\two]\right)\ec
\ee
with
\be\label{2d bZ TQFT}
    \cT_{2d}^{p}[A\one,f\two]=\frac{i}{2\pi}\int_{\Sigma\two} \phi\one\wedge d\Upsilon\zero + \Upsilon\zero\wedge f\two +p\phi\one \wedge A\one\ec
\ee
where $\Upsilon\zero$ is a $U(1)$-valued compact scalar field while $\phi\one$ is a $\bR$-valued 1-form gauge field. The above theory describes a 2d mixed $U(1)/\mathbb{R}$ BF theory coupled to the bulk $U(1)$ gauge field $A\one$ and the bulk $\bR$-valued 2-form $f\two$.

This operator \eqref{genT} is gauge invariant if we consider the following gauge transformations for $\phi\one$ and $\Upsilon\zero$:
\be
\begin{split}
    &A\one\rightarrow A\one +d\lambda_A\zero\ec \quad f\two\rightarrow f\two+d\lambda_f\one \ec \quad\\
    &\Upsilon\zero\rightarrow \Upsilon\zero+2\pi n_{\Upsilon}-2m\lambda\zero_A\ec \quad  \phi\one\rightarrow \phi\one+d\lambda_{\phi}\zero -\ \lambda_f\one \ed
\end{split}
\ee
where $n_{\Upsilon}\in \bZ$. Note that, given the above transformation rules, we can define non-genuine topological operators of the 2d TQFT $\cT_{2d}^{p}[A\one, f\two]$:
\begin{equation} \label{junctions T}
\exp\left(i\beta\int_{\gamma\one} \phi\one+i\beta \int_{\Sigma\two_{\gamma}} f\two\right)\ec\quad \exp\left(iq\left(\Upsilon\zero(x_f)-\Upsilon\zero(x_i)+2m\int_{x_i}^{x_f}A\one\right)\right)\ec
\end{equation}
with $\partial\Sigma\two_{\gamma}=\gamma\one$ and where the open surface and line, respectively, typically lie outside the surface where the 2d TQFT is defined. These operators then allow the $U_{\beta}$ and $W_{2m n}$ defects \eqref{U,W} to end topologically on $T_m$ for any $\beta\in \bR$ and $n\in \bZ$. This is necessary for consistency with \eqref{5d VT} and \eqref{5d TT}. Note also that when $T_m$ ends at the boundary,  we must consider boundary conditions that are consistent with the fact that $U_{\beta}$ cannot terminate on $\mathcal{M}$ while $W_{2m n}$ can always do that. Therefore, we consider Dirichlet boundary conditions for $\Upsilon\zero$ and Neumann boundary conditions for $\phi\one$.

We would now also like to recover the non-invertible $U(1)_A$ symmetry defect $\cD_\alpha(\Pi\three)$ of the physical theory that we introduced in section \ref{NoninvU1}. Naturally, they should be related to the bulk operators $V_{\alpha}(\Pi\three, \Omega\four)$ defined in \eqref{nongenV}. Since the former are genuine, i.e.~they depend only on the closed 3-surface $\Pi\three$ on which they are defined, we then need to remove the dependence on the open surface $\Omega\four$ from the operator $V_{\alpha}(\Pi\three, \Omega\four)$.\footnote{Indeed, the integral over $\Omega\four$ involves the field $f_2$ on which we have to impose Neumann boundary conditions, and hence does not trivialize on the boundary.} 

Unsurprisingly, this can be achieved by stacking the operator \eqref{nongenV} with the mixed $U(1)/\bR$ TQFT $\cT^\alpha$ introduced in \eqref{Talpha}:
\be\label{genV}
V_{\alpha}(\Pi\three)=\exp\left(\frac{i\alpha}{4\pi}\int_{\Pi\three}b\three + \cT^{\alpha}[f\two]\right)\ed
\ee
Note that in the present symmetry TFT context, the 3d TQFT $\cT^{\alpha}$ is coupled to $f\two$ which is  the bulk equivalent of the $U(1)\one$ magnetic symmetry current of the physical theory. 

Using \eqref{gaugeTrans4dTFT} and the fact that the fields $c\one$ and $\Phi\one$ defining $\cT^\alpha$ also transform under a gauge transformation of $f\two$, as
\begin{equation}
c\one\to c\one+d\lambda_c\zero -\lambda_f\one\ec\qquad
\Phi\one\to\Phi\one +d\lambda_\Phi\zero-\frac{\alpha}{2\pi}\lambda_f\one\ec
\end{equation}
it is easy to see that $V_{\alpha}(\Pi\three)$ is gauge-invariant. Furthermore, the bulk $U_\beta$ operators can end on the lines of the $\cT^\alpha$ theory:
\begin{equation} \label{junctions V}
\exp\left(i\beta\int_{\gamma\one} c\one+i\beta \int_{\Sigma\two_{\gamma}} f\two\right)\ec\quad \exp\left(iq\int_{\gamma\one}\Phi\one+i\frac{q\alpha}{2\pi}\int_{\Sigma\two_{\gamma}}f\two\right)\ec
\end{equation}
where again $\Sigma\two_\gamma$ is such that $\partial\Sigma\two_{\gamma}=\gamma\one$ and it typically extends outside $\Pi\three$.

From the definition of genuine operators \eqref{genT} and \eqref{genV} it is also possible to study how they act on each other and reproduce the relations \eqref{5d VT} and \eqref{5d TT}. The main difference here is that \( U_{m\alpha} \) and \( W_{2nm} \) now terminate on the junction operators introduced in \eqref{junctions T} and \eqref{junctions V}. The operator \( U_{m\alpha} \) terminates on \( V_{\alpha} \) with the line operator $\exp(im\int \Phi\one)$ and on  \(T_m\) with \(\exp(i\frac{\alpha m}{2\pi}\int \phi\one)\). At the same time the operator  $W_{2mn}$ terminates on $T_m$ with a local operator \( \exp(in\Upsilon\zero) \) and similarly for $T_n$. This ensures the topological nature of these junctions.

When $T_m$ terminates on the boundary $\cM_P$, and $V_\alpha$ is pushed there, the relation \eqref{5d VT} reproduces the non-invertible action on 't Hooft lines given in \eqref{nongenHq}.

The operators $T_m(\Sigma\two)$ and $V_\alpha(\Pi\three)$ presented here differ from the ones discussed in \cite{Antinucci:2024zjp}.\footnote{If one considers the 5d theory \eqref{symQED} on a closed manifold, then all topological operators are symmetry defects. It is then straightforward to see that the defects $T_m$ and $V_\alpha$, as defined in \eqref{genT} and \eqref{genV} respectively, generate non-invertible 2-form and 1-form symmetries taking values in $\bZ$ and $U(1)$.} Similarly to the discussion of section \ref{NoninvU1}, a crucial role in this context is played by the dressing of topological defects with mixed $U(1)/\bR$ theories $\cT_{2d}^{p}[A\one, f\two]$ and $\cT^{\alpha}[f\two]$. To see the difference, let us for example consider the non-genuine operator \eqref{nongenV}. As discussed in \cite{Antinucci:2024zjp}, when $\alpha  = 2\pi p/N \in 2\pi\mathbb{Q}$ with $\gcd(p,N)=1$, one can obtain a genuine topological operator by dressing \eqref{nongenV} with $\cA^{N,p}[f\two]$. The drawback with this choice of dressing is that all the remaining topological operators \eqref{nongenV} for irrational values of $\alpha$ remain non-genuine. Our proposal is that such issue is circumvented by the 3d mixed $U(1)/\bR$ theory $\cT^\alpha$ which allows us to define genuine topological operators for any value of $\alpha \in [0,2\pi)$. In the case of \eqref{nongenT}, $T_m$ could become genuine for any $m\in \bZ$ using only $U(1)$-valued fields. However, using only $U(1)$ fields is inconsistent with \eqref{5d VT}, if we allow $\alpha$ to take any values in $[0,2\pi)$.

\section*{Acknowledgements}\noindent
We thank Jeremias Aguilera Damia, Andrea Antinucci, Pietro Benetti Genolini, Francesco Benini, Giovanni Galati and Elise Paznokas for helpful discussions and comments on the draft. AA, RA and LT are respectively a Research Fellow, a Research Director and a Postdoctoral Researcher of the F.R.S.-FNRS (Belgium). This research is further supported by IISN-Belgium (convention 4.4503.15) and through an ARC advanced project.

\appendix

\section{Non-Invertible Fusion Rules and Comparison with $\cA^{N, p}$ Theory}\label{fusionsection}

In section \ref{condensection}, we studied the parallel fusion of $\cD_\alpha(\Pi\three)$ with its orientation reversal. We now consider the more general parallel fusion of $\cD_{\alpha}(\Pi\three)$ and $\cD_{\beta}(\Pi\three)$ with $\alpha, \beta \in [0,2\pi)$:
\be
\begin{split}
    &\cD_{\alpha}(\Pi\three)\times \cD_{\beta}(\Pi\three)=\exp\left(\frac{i}{2}(\alpha+\beta)\int_{\Pi\three}* j\one_A\right)\int D[c\one,\Phi\one,\overline{c}\one,\overline{\Phi}\one]\\
    &\exp\left(\frac{i}{2\pi}\int_{\Pi\three}-\frac{\alpha}{4\pi}c\one \wedge dc\one +\Phi\one \wedge (dc\one+dA\one)-\frac{\beta}{4\pi}\overline{c}\one \wedge  d\overline{c}\one+\overline{\Phi}\wedge \one( d\overline{c}\one+dA\one)\right)\ed
\end{split}
\ee
After performing the following field redefinition,
\be
\Phi\rightarrow\Phi -\overline{\Phi}\ec \quad \overline{c}\one\rightarrow \overline{c}\one+c\one \ec\quad \overline{\Phi}\one\rightarrow \overline{\Phi}\one+\frac{\beta}{2\pi} c\one+\frac{\beta}{4\pi}\overline{c}\one\ec
\ee
we get:
\be
\begin{split}
    &\cD_{\alpha}(\Pi\three)\times \cD_{\beta}(\Pi\three)=\exp\left(\frac{i}{2}(\alpha+\beta)\int_{\Pi\three}* j\one_A\right)\int D[\overline{c}\one,\overline{\Phi}\one]\exp\left(\frac{i}{2\pi}\int_{\Pi\three}\overline{\Phi}\one \wedge d\overline{c}\one\right)\times\\
    &\int D[c\one,\Phi\one]\exp\left(\frac{i}{2\pi}\int_{\Pi\three}-\frac{\alpha+\beta}{4\pi}c\one\wedge dc\one+\Phi\one \wedge(dc\one+dA\one)\right)\ed
\end{split}
\ee
The first path integral is the normalization factor \eqref{normalization}, already obtained in section \ref{condensection}. The other terms correspond to the definition of $\cD_{\alpha+\beta}(\Pi\three)$. We conclude
\be
\cD_{\alpha}(\Pi\three)\times \cD_{\beta}(\Pi\three)=(\mathcal{Z}_{U(1)/\bR})\ \cD_{\alpha+\beta}(\Pi\three)\ed
\ee
Notice that the above relation is consistent with \eqref{condensation U(1)U(1)} since $\cD_{0}(\Pi\three)=\cC_{U(1)}(\Pi\three)$.

We now consider the fusion of continuous non-invertible operators with the operators of the non-invertible $2\pi\mathbb{Q}/\bZ$ symmetry defined in \cite{Choi:2022jqy}.\footnote{See \cite{Putrov:2022pua} for a recent discussion on QFT with a $\mathbb{Q}/\bZ$ symmetry.} For this, we study the line operators of the TQFT stacked with those operators.

The $\mathcal{A}^{N,p}$ theory of \cite{Hsin:2018vcg} is a 3-dimensional TQFT with $\bZ_N\one$ symmetry and genuine line operators that we denote by $V(\gamma)$. The parameter $p\in\bZ_N$ satisfies $\gcd(N,p)=1$. Those line operators braid as:
\begin{equation}
    \langle V(\gamma\one_1)V(\gamma\one_2)\rangle =\exp\left(2\pi i\frac{p^{-1}_N}{N}\,\text{Link}(\gamma\one_1,\gamma\one_2)\right)\ec
\end{equation}
where $p^{-1}_N$ is the inverse modulo $N$ of $p$. When the theory is coupled to a background gauge field $B\two$, the operators $V$ become non-genuine line operators. In particular, when $B\two=\frac{dA\one}{N}$, the non-genuine line operators of the $\mathcal{A}^{N,p}[dA\one/N]$ theory are given by
\begin{equation}
    V[A\one]=V[0]\exp\left(\frac{i}{N}\int_{\Sigma_{\gamma}\two}dA\one\right)\ec
\end{equation}
where $\Sigma_{\gamma}\two$ satisfies $\partial\Sigma_{\gamma}\two=\gamma\one$.

It is easier to study the line operators of the theory $\cT^{\alpha}[dA\one]$ \eqref{Talpha} with the following alternative expression
\be\label{Talphahat}
\cT^\alpha[dA\one] = \frac{i}{2\pi}\int_{\Pi\three}\hat\Phi\one \wedge (dc\one+dA\one)+\frac{\alpha}{4\pi}c\one \wedge dA\one\ec
\ee
where $\hat\Phi\one=\Phi\one-\frac{\alpha}{4\pi}c\one$, and it has the same quantized fluxes and gauge transformations as $\Phi\one$. The line operators are then
\begin{equation}
\begin{split}
    U^{\theta}(\gamma\one)[A\one]=\exp\left(i\frac{\theta}{2\pi}\int_{\gamma\one}c\one+i\frac{\theta}{2\pi}\int_{\Sigma_{\gamma}\two}dA\one\right)\ec\\ \quad W^q(\gamma\one)[A\one]=\exp\left(iq\int_{\gamma\one}\hat\Phi\one+iq\frac{\alpha}{4\pi}\int_{\Sigma_{\gamma}\one}dA\one\right)\ec
    \end{split}
\end{equation}
with $\theta\in \bR/(2\pi\bZ)$ and $q\in \bZ$. Those operators braid with each other as:
\begin{equation}
    \langle U^{\theta}(\gamma\one_1) W^q(\gamma\one_2)\rangle = \exp\left(i\theta q\,\text{Link}(\gamma\one_1,\gamma\one_2)\right)\ed 
\end{equation}
Note that if we define $\hat W=WU^\pi$, we observe that it still braids trivially with itself, and braids with $U^\theta$ exactly as $W$. However, $\hat W[A\one]=\hat W[0]\exp(i\frac{\alpha+2\pi}{4\pi}\int dA\one)$, therefore those operators define the theory $\cT^{\alpha+2\pi}$. We conclude that  $\mathcal{T}^{\alpha}$ and $\mathcal{T}^{\alpha+2\pi}$ have the same operators and therefore correspond to the same theory.

Using the line operators of both theories, we can define the following line operators:
\begin{equation}
    \tilde{V}[A\one]=V[A\one]U^{-2\pi /N}[A\one]\ec \qquad \tilde{W}[A\one]=W[A\one]V^{p}[A\one]U^{\frac{-2\pi p}{2N}}[A\one]\ed 
\end{equation}
These new line operators, together with $U^{\theta}$, also form a basis of the operator content of the $\cA^{N,p}[dA\one/N]\times \cT^{\alpha}[dA\one]$ theory. 
Now, the lines $\tilde{V}$ are independent of $A\one$,  have the same self-braiding as the $V$ lines, and braid trivially with $\tilde W$ and $U^\theta$.  They therefore define a (decoupled) $\cA^{N,p}[0]$ theory. The line $\tilde{W}$ also still braids with $U^\theta$ in the same way as $W$, while its dependence on the background field $A\one$ is given by:
\begin{equation}
    \tilde{W}^q[A\one]=\tilde{W}^q[0]\exp\left(iq\frac{1}{4\pi}\left(\alpha+\frac{2\pi p}{N}\right)\int_{\Sigma_{\gamma}\two}dA\one\right)\ed    
\end{equation}
Therefore, they define a $\cT^{\alpha+\frac{2\pi p}{N}}[dA\one]$ theory, and
we conclude that 
\begin{equation}
    \cA^{N,p}[dA\one/N]\times \cT^{\alpha}[dA\one]=\cA^{N,p}[0]\times \cT^{\alpha+\frac{2\pi p}{N}}[dA\one]\ed
\end{equation}
This relation can be used to obtain the fusion rule of $\cD_{\alpha}$ and $\cD^{\mathbb{Q}}_{2\pi p/N}$:
\begin{equation}
    \cD_{\alpha}(\Pi\three)\times \cD^{\mathbb{Q}}_{2\pi p/N}(\Pi\three)=(\CZ_{\cA^{N,p}})\ \cD_{\alpha+2\pi p/N}(\Pi\three)\ec
\end{equation}
where $\CZ_{\cA^{N,p}}$ is the partition function of the minimal 3d TQFT $\cA^{N,p}$ evaluated on $\Pi\three$.
In particular, by setting $\alpha=0$, we obtain a relation between operators defined with the $\cT^{\frac{2\pi p}{2N}}[dA\one]$ theory and the $\cA^{N,p}[dA\one/N]$ theory:
\begin{equation}
    (\CZ_{\cA^{N,p}})\ \cD_{2\pi p/N}(\Pi\three)=\cC_{U(1)}(\Pi\three)\times \cD^{\mathbb{Q}}_{2\pi p/N}(\Pi\three)\ed
\end{equation}
We conclude that when $\alpha\in 2\pi\mathbb{Q}$, the operator $\cD_{\alpha}(\Pi\three)$ is not a minimal operator.

\section{General Remarks on 3d TQFTs with $U(1)$ and $\bR$ Gauge Fields}

The action \eqref{Talpha} for $\cT^{\alpha}[B\two]$ is invariant under rescalings of $c\one$ by any real number.
We can then rescale by $N$ and then take $\alpha=2\pi p/N$, with $N,p\in\bZ$ satisfying $\gcd(p,N)=1$. In this case, the action is well defined even if $c\one$ is promoted to be a $U(1)$-valued field. If we replace $c\one$ by $C\one$, we then get
\be \label{explicit ANp}
\frac{i}{2\pi}\int_{\Pi\three}-\frac{Np}{2}C\one \wedge dC\one+\Phi\one \wedge (N dC\one+B\two)\ed
\ee
This is the action of the theory $\mathcal{A}^{N,p}[B\two/N]\times \mathcal{A}^{N,-p}[0]$, which was considered in appendix $C$ of \cite{Choi:2022jqy} as an example of a non-minimal theory to define operators implementing transformations of the non-invertible $2\pi \mathbb{Q}/\bZ$ axial symmetry. As for \eqref{Talpha}, this action can be obtained by two successive topological gaugings in half spacetime. In this case, the gauged symmetries are a $\bZ_N\one$ subgroup of the $U(1)\one$ magnetic symmetry and its Pontryagin dual $\hat{\bZ}_N\one$ symmetry.

More generally, to obtain a 3d theory with an anomaly $-\frac{i\alpha}{8\pi^2}\int_{\Omega}B\two\wedge B\two$, we can consider an action of the form:
\be\label{anomalous higher gauging}
\frac{i}{2\pi}\int_{\partial \Omega\four}-\frac{\alpha}{4\pi}\psi\one \wedge d\psi\one+\phi\one \wedge ( d\psi\one+B\two)\ec
\ee
where we can choose $\psi\one$ and $\phi\one$ to either be $\bR$-valued or (proportional to) $U(1)$-valued fields. The symmetries of the theory and the allowed values of $\alpha$ depend on this choice. When $\alpha=0$, this theory gives a condensation defect on $\partial\Omega\four$, which corresponds to topologically gauging a symmetry whose current is $(* j)\two=\frac{1}{2\pi}B\two$. Setting $\alpha\neq 0$ corresponds to gauging this symmetry with torsion. Operators defined using \eqref{anomalous higher gauging} act non-invertibly on operators charged under this symmetry.

The type of gauged symmetry depends on the choice of fields $\psi\one$ and $\phi\one$:
\begin{enumerate}
    \item If $\psi\one=NC\one$ and $\phi\one=\Phi\one$, where $N\in\bZ$ and $C\one$ and $\Phi\one$ are $U(1)$ valued fields, we recover \eqref{explicit ANp}. In this case, we higher-gauge in $\partial\Omega\four$ a $\bZ_N$ symmetry. This theory is only well defined if $\int B\in 2\pi \bZ$ and $\alpha\in \frac{2\pi}{N}\bZ$. The theory is not a minimal $\bZ_N$ theory and can be expressed as $\mathcal{A}^{N,p}[B\two/N]\times \mathcal{A}^{N,-p}[0]$.
    \item If $\psi\one$ is $\bR$ valued and $\phi\one$ is $U(1)$ valued, we obtain the theory $\cT^{\alpha}[B\two]$ for any $\alpha$. In this case, we higher-gauge a $U(1)$ symmetry. This theory is only well defined if $\int B\in 2\pi \bZ$. 
    \item If $\psi\one$ is $U(1)$ valued and $\phi\one$ is $\bR$ valued, we higher-gauge a $\bZ$ symmetry. This theory is only well defined if $\alpha\in 2\pi\bZ$. 
    \item If $\psi\one$ and $\phi\one$ are both $\bR$ valued, we higher-gauge a $\bR$ symmetry. This theory is not a minimal $\bR$ theory. Indeed, 
    performing the field redefinition $\psi\one\to\psi\one+\frac{2\pi}{\alpha}\phi\one$ one obtains two decoupled Chern-Simons-like $\bR$ theories, with $B\two$ coupling only to $\phi\one$. This implies that a single dynamical field is enough to define a theory with symmetry $\bR$ and anomaly $-\frac{i\alpha}{8\pi^2}\int_{\Omega}B\two\wedge B\two$. This minimal $\bR$ theory is considered in appendix \ref{minimalRsection}. 
\end{enumerate}
Note that both $\bR$ and $\bZ_N$ are self-dual under Pontryagin duality. Therefore, a single $\bR$-valued field is enough to define a minimal theory with a $\bR$ symmetry, and a single $\bZ_N$-valued field is enough to define a minimal theory with a $\bZ_N$ symmetry. In this case, the charged operators are also the symmetry operators of the theory. 

However, the groups $U(1)$ and $\bZ$ are exchanged under Pontryagin duality. In this case, an operator cannot be simultaneously a charged object and symmetry operator of the same symmetry. To define theories with those symmetries, we need at least one $\bR$-valued and one $U(1)$-valued field. Therefore, the theory \eqref{anomalous higher gauging} is a minimal theory only when we consider a $U(1)$ or $\bZ$ symmetry but it is not minimal when we consider a $\bR$ or a $\bZ_N$ symmetry.

\section{Non-Invertible Defects with Minimal $\bR$-TQFT}\label{minimalRsection}
Let us consider the following theory:
\be\label{Ralpha}
\bR^\alpha[B\two] = \frac{i}{2\pi}\int_{\Pi\three}\frac{\pi}{\alpha}c\one\wedge dc\one+c\one\wedge B\two\ed
\ee
where $c\one$ is a dynamical $\bR$-valued 1-form gauge field. The above TQFT is nothing but 3d Chern-Simons theory with gauge group $\bR$ coupled a background field $B\two$ for its $\bR\one$ 1-form  symmetry.\footnote{This theory was previously explored in a different setting in \cite{Maloney:2020nni,Benini:2022hzx}.} This is the minimal 3-dimensional TQFT with  $\bR\one$ symmetry and anomaly $-\frac{i\alpha}{8\pi^2}\int_{\Omega}B\two\wedge B\two$. The $\bR^\alpha[B\two]$ theory can be defined as the boundary theory of the following 4-dimensional TQFT:
\be
S[B\two]=-\frac{i}{2\pi}\int_{\Omega\four}\frac{2\pi}{\alpha} b\two \wedge dc\one+\frac{\pi}{\alpha} b\two\wedge b\two+ b\two\wedge B\two\ec
\ee
where $b\two$, $c\one$ are $\bR$-valued dynamical gauge fields. When $\Omega\four$ is closed, the theory is invariant under the following gauge transformations:
\be
b\two\rightarrow b\two+d\lambda_b\one\ec \qquad c\one\rightarrow c\one+d\lambda_c\zero-\lambda_b\one\ed
\ee
Since $c\one$ and $\lambda_b\one$ are $\bR$-valued gauge field, we have $\int_{\Sigma\two} d\lambda_b\one=0=\int_{\Sigma\two} dc\one$ for any closed sub-manifold $\Sigma\two\subset\Omega\four$. 
This action can be interpreted as the topological gauging of an $\bR$ symmetry whose current is $(* j)\two=\frac{1}{2\pi}B\two$, with torsion.
When $\Omega\four$ is open, we consider Dirichlet boundary conditions for $b\two$. After integrating out $b\two$, we get back \eqref{Ralpha} on $\partial\Omega\four$ together with the compensating anomaly term.

The theory $\bR^{\alpha}[dA\one]$ has the right anomaly to cancel the non-gauge invariant part of an operator implementing an axial transformation of parameter $\alpha$. We can consider the following operator:
\be\label{newnoninv}
\tilde\cD_{\alpha}(\Pi\three)= \exp\left(i\int_{\Pi\three}\frac{\alpha}{2}*j\one_A + \bR^\alpha[dA\one]\right)\ed
\ee
This operator is topological and gauge invariant. However, when defining the $\bR^{\alpha}[dA\two]$ theory, we ignored the quantization $\int dA\one\in 2\pi\bZ$. This means that since $* \frac{1}{2\pi}dA\one$ is the current of a $U(1)\one$ symmetry, the higher-gauged $\bR$-symmetry is a non-faithful symmetry. This is the main reason why we prefer to use the mixed $U(1)/\bR$ theory $\cT^\alpha$ to build our non-invertible axial symmetry defects.

It is nevertheless instructive to see how we would be able to (partially) reproduce our results with this definition of the non-invertible defects. 
As done for the operators defined with the $\cT^{\alpha}[dA\one]$ theory, we can start by seeing how \eqref{newnoninv} acts non-invertibly on operators charged under the $U(1)\one$ magnetic symmetry. The equations of motion of the $\bR^{\alpha}$ theory impose
$\int_{\Sigma}dA\one=-{\frac{2\pi}{\alpha}}\int_{\Sigma}dc\one=0$. Let us now consider the insertion of an operator 
\be\label{operator Ralpha}
\exp\left(iq\int_{\gamma}c\one+i\frac{q\alpha}{2\pi}\int_{\Sigma_\gamma\two} dA\one\right)\ec
\ee with $\gamma\one\subset \Pi\three$ and $\partial\Sigma_{\gamma}\two=\gamma\one$. In this case, the equations of motion imply
\be \label{Ralpha flux qunatization}
\int_{\Sigma\two}dA\one=-2\pi q I(\Sigma\two,\gamma\one)\ed
\ee
Since the operator \eqref{operator Ralpha} is well defined for any $q\in \bR$, this condition is too lax with respect to the quantization $\int_{\Sigma}dA\one\in 2\pi\bZ$. Only lines with $q\in\bZ$ need to be inserted when \eqref{newnoninv} crosses a magnetic operator $H_q$. The latter then becomes non-genuine and gets attached to a magnetic symmetry surface:
\be
H_q\exp\left(i\frac{q\alpha}{2\pi}\int_{\Sigma_\gamma\two}dA\one\right)\ec
\ee
which is the same as the one obtained when considering the non-invertible operator defined with the $\cT^{\alpha}$ theory.

We can further consider the condensation defect obtained through the parallel fusion of $\tilde\cD_\alpha$ with  $\tilde{\cD}_{-\alpha}$:
\be
\tilde\cD_{\alpha}(\Pi\three)\times\tilde{\cD}_{-\alpha}(\Pi\three) = \exp\left(
\frac{i}{2\pi}\int_{\Pi\three}\frac{\pi}{\alpha}c\one\wedge dc\one-\frac{\pi}{\alpha}\bar c\one \wedge d\bar c\one +(c\one+\bar c\one)\wedge dA\one\right)\ed
\ee
After performing shifts of $c\one$ and $\bar c\one$ and a final rescaling of $\bar c\one$, the right hand side of the above expression can be rewritten as
\be
\cC_{\bR}(\Pi\three) = \exp\left(
    \frac{i}{2\pi}\int_{\Pi\three} c\one\wedge d\overline{c}\one+ c\one\wedge dA\one\right)\ed
\ee
The topological operator $\cC_{\bR}(\Pi\three)$ should be interpreted as a condensation defect for a $\bR$-symmetry. Integrating out $\overline{c}\one$ constrains $c\one$ to be a closed form. The path integral is then over $H^1(\Pi\three, \bR)$. The condensation defect can therefore be written as
\be
\begin{split}\label{condR}
    \cC_{\bR}(\Pi\three)&= \prod_{\gamma_i\two\in X}\left[\int_{-\infty}^{\infty}ds_i\exp\left(\frac{is_i}{2\pi}\int_{\gamma_i\two}dA\one\right)\right]\ec
\end{split}
\ee
where $X$ is a minimal generating set of $H_2(\Pi\three,\bZ)$.
This corresponds to topologically higher-gauging a $\bR$ symmetry on $\Pi\three$. Since we have not taken into account the quantization of $\int dA\one$, we obtain an infinite number of copies of the condensation defect $\cC_{U(1)}$. 

Finally, the parallel fusion of two operators \eqref{newnoninv} with two different parameters $\alpha$ and $\beta$ gives
\be
\tilde\cD_{\alpha}(\Pi\three)\times\tilde\cD_{\beta}(\Pi\three)=\exp\left(i\int_{\Pi\three}\frac{(\alpha+\beta)}{2}* j_A\one+\bR^{\alpha}[dA\one]+\bR^{\beta}[dA\one]\right)\ed
\ee
The non-trivial part of this fusion is given by
\begin{align*}
   &\int D[c\one,\overline{c}\one]\exp\left(\frac{i}{2\pi}\int\left(\frac{\pi}{\alpha}c\one\wedge dc\one +  c\one\wedge dA\one+\frac{\pi}{\beta}\overline{c}\one\wedge d\overline{c}\one + \overline{c}\one\wedge dA\one\right)\right)\ed
\end{align*}
With a sequence of shifts and rescalings (and assuming $\alpha+\beta\neq 0$), the path integrals become
\begin{align*}
    \int D[\bar c\one]\exp&\left(\frac{i}{4\pi}\int_{\Pi\three}\bar c\one \wedge d\bar c\one\right)
    \times\\ &
    \int D[c\one]\exp\left(\frac{i}{2\pi}\int_{\Pi\three}\frac{\pi}{\alpha+\beta}c\one \wedge dc\one+c\one\wedge dA\one\right)\ed
\end{align*}
It follows that
\be
\tilde\cD_{\alpha}(\Pi\three)\times\tilde\cD_{\beta}(\Pi\three) = (\mathcal{Z}_{\bR})\ \tilde\cD_{\alpha+\beta}(\Pi\three) \ec
\ee
where $\mathcal{Z}_{\bR}$ is the partition function of Chern-Simons theory with $\bR$ gauge group on $\Pi\three$. To better understand the relation between $\cD_{\alpha}$ and $\tilde{\cD}_{\beta}$, let us now consider the fusion between these two operators:
\be
\tilde{\cD}_{\alpha}(\Pi\three)\times \cD_{\beta}(\Pi\three)=\exp\left(i\int_{\Pi\three}\frac{(\alpha+\beta)}{2}* j_A\one+\bR^{\alpha}[dA\one]+\cT^{\beta}[dA\one]\right)\ed
\ee
This expression contains the path integral
\begin{align*}
    \int D[\overline{c}\one]\exp &\left(\frac{\pi}{\alpha}\overline{c}\one\wedge d\overline{c}\one + \overline{c}\one\wedge dA\one\right)\times \\
    &\int D[\Phi\one,c\one] \exp \left(\frac{i}{2\pi}\int-\frac{\beta}{4\pi}c\one\wedge dc\one + \Phi\one\wedge( dc\one + dA\one)\right)\ec
\end{align*}
which, with some fields redefinitions, can be written as
\begin{align*}
    &\int D[\Phi\one,c\one,\overline{c}\one]\exp\left(\frac{i}{2\pi}\int\left(\frac{\pi}{\alpha}\overline{c}\one\wedge d\overline{c}\one-\frac{\alpha+\beta}{4\pi}c\one\wedge dc\one +\Phi\one\wedge (dc\one +dA\one)\right)\right)\ed
\end{align*}
The first term is the action of a Chern-Simons theory with a $\bR$-valued field and the other ones define the $\cT^{\alpha+\beta}[dA\one]$ theory. The fusion rule can therefore be written as 
\be
\tilde{\cD}_\alpha(\Pi\three)\times \cD_{\beta}(\Pi\three)=(\mathcal{Z}_{\bR})\ \mathcal{D}_{\alpha+\beta}(\Pi\three)\ed
\ee
Setting $\beta=0$, we conclude
\be
\mathcal{D}_{\alpha}(\Pi\three)=\frac{1}{\mathcal{Z}_{\bR}}\,\cC_{U(1)}(\Pi\three)\times\tilde{\cD}_\alpha(\Pi\three)\ed
\ee
For any $\alpha\neq 0$, the operator $\mathcal{D}_{\alpha}(\Pi\three)$ can always be expressed as the fusion of the operator $\tilde\cD_{\alpha}(\Pi\three)$ with $\cC_{U(1)}(\Pi\three)$. The condensation defect $\cC_{U(1)}$ contains the information about the compact nature of the $U(1)$ magnetic symmetry.
Note however that $\cC_{U(1)}$ and $\tilde{\cD}_\alpha$ act on $H_q$ in the same (non-invertible) way, therefore $\cD_\alpha$ does not annihilate more operators than $\tilde{\cD}_\alpha$ despite being a less minimal operator.

We thus conclude that the price to pay for this seemingly more minimal presentation of the non-invertible defects is that the compact structure of the $U(1)$ symmetry groups does not come out naturally, but has to be introduced as an external input.

\bibliographystyle{ytphys}
\baselineskip=0.85\baselineskip
\bibliography{HigherGaugings}

\end{document}